\documentstyle[emulateapj,apjfonts]{article}
\begin{document}
\newcommand{\C}{$^{12}{\rm C} \ $}
\newcommand{\Ru}{$^{104}{\rm Ru} \ $}

\submitted{Submitted to The Astrophysical Journal}

\title{Carbon Flashes in the Heavy Element Ocean on Accreting Neutron
Stars}

\author{Andrew Cumming and Lars Bildsten }
\affil{Institute for Theoretical Physics\\
Kohn Hall, University of California, Santa Barbara, CA 93106 \\
cumming@itp.ucsb.edu, bildsten@itp.ucsb.edu}

\begin{abstract} 
We show that the burning of a small mass fraction $X_{12}$ of \C in a
neutron star ocean is thermally unstable at low accumulated masses
when the ocean contains heavy ashes from the hydrogen burning rapid
proton (rp) process. The key to early unstable ignition is the
decreased thermal conductivity of a heavy element ocean. The
instability requires accretion rates, $\dot M$, in excess of one-tenth
the Eddington limit when $X_{12}< 0.1$. Lower $\dot M$'s will stably
burn a small mass fraction of \C. The unstable flashes release $\sim
10^{42}-10^{43}{\rm ergs}$ over hours to days and are likely the cause
of the recently discovered large Type I bursts (so-called
``superbursts'') from six Galactic low-mass X-ray binaries. In
addition to explaining the energetics, recurrence times and durations
of the superbursts, these mixed \C flashes also have an $\dot M$
dependence of unstable burning similar to that observed. Though the
instability is present at accretion rates $\approx \dot M_{\rm Edd}$,
the flashes provide less of a contrast with the accretion luminosity
there, thus explaining why most detections are made at $\dot M\approx
(0.1-0.3)\dot M_{\rm Edd}$. Future comparisons of time dependent
theoretical calculations with observations will provide new insights
on the rp-process.

\end{abstract}

\section{Introduction}\label{sec:Intro}

The monitoring of galactic sources by the {\it Wide-Field Camera} on
BeppoSAX and the {\it All-Sky Monitor} on the Rossi X-Ray Timing
Explorer have detected large Type I bursts (hereafter referred to as
``superbursts'') from six accreting neutron stars (Cornelisse et
al. 2000; Strohmayer 2000; Heise et al. 2000; Wijnands 2001; Kuulkers
2001). These bursts share a number of characteristics; burst energies
$\approx 10^{42}\ {\rm erg}$ (roughly 1000 times larger than a normal
Type I burst); durations of a few hours and accretion rates $\approx
(0.1-0.3)\dot M_{\rm Edd}$ (Wijnands 2001). The superburst recurrence
time is not well known, though one recurrent burst has been seen 4.7
years later (Wijnands 2001). The light curves are similar to a Type I
burst in spectral evolution, though the immediate decay timescale is
about two hours; 1000 times longer than a typical Type I X-ray
burst. The superbursts appear to be thermonuclear flashes from fuel at
much larger depths than a typical Type I burst.  For example, if the
nuclear energy release is 1 MeV per accreted nucleon ($E_{\rm
nuc}=10^{18} {\rm erg \ g^{-1}}$), the accumulated mass would be
$10^{24} \ {\rm g}$, implying a recurrence time of $\approx 4$ months
for an accretion rate of $\dot M\approx 0.1 \dot M_{\rm Edd}\approx
10^{17} \ {\rm g \ s^{-1}}$, much longer than a typical Type I burst.

Flashes from pure \C layers have been discussed previously (Woosley \&
Taam 1976; Taam \& Picklum 1978; Brown \& Bildsten 1998) and might
well apply to the pure helium accretor 4U 1820-30 (Strohmayer
2000). However, there are two difficulties with such a model for the
superbursts from hydrogen/helium accretors: (1) the recurrence times
are very long, and the burst energies correspondingly large, unless a
large heat flux from the crust of the star is present deep in the
ocean; and, (2) our current understanding of the ashes from
hydrogen/helium burning does not point to accumulation of pure
\C. Instead, Schatz et al. (1999, 2001) have shown that only a small
amount of \C remains after all of the hydrogen and helium has burned
via the rapid proton (rp) process (Wallace \& Woosley 1981); either
from steady state burning or unstable type I X-ray bursting. Even
though a small fraction by mass (typically $X_{12}\approx 0.05-0.1$),
we show here that it is enough to trigger a thermonuclear runaway with
energy comparable to the superbursts. Most important, however, is the
role played by the heavy ashes from the rp process. These reduce the
thermal conductivity enough to force a large temperature gradient in
the ocean. This ignites the \C at much lower masses than previously
expected and reduces the energy and recurrence times to levels
consistent with the superbursts. In addition, the conductive cooling
times from these flashes are consistent with the long decay times
observed.

\section{Carbon Ignition in a Heavy Element Ocean}

We presume that the ashes of hydrogen/helium burning in the upper
atmosphere primarily consist (by mass) of a single dominant nucleus of
mass $Am_p$ and charge $Ze$ and a small amount of \C with mass
fraction $X_{12}$. Schatz et al.~(2001) recently showed that the onset
of a closed cycle (called the SnSbTe cycle) in the rp-process
naturally stops the increasing $A$ trajectory once nuclei of mass
$A\approx 104$ are reached. Though a range of nuclei are made, those
with masses in this region dominate the mix. For simplicity, we take
\Ru as our fiducial nucleus ($A=104$ and $Z=44$) with mass fraction
$1-X_{12}$.

The energy released when \C fuses (for simplicity, we presume that
the reaction makes $^{24} {\rm Mg}$) is $E_{\rm nuc}\approx 5.6\times
10^{17}\ {\rm erg \ g^{-1}}$, so that if $X_{12}=0.05$, an accumulated
mass of $3.6\times 10^{25} {\rm g}$ is needed for a $10^{42} \ {\rm
erg}$ burst. At an accretion rate of $0.1\ \dot m_{\rm Edd}$ onto a
$10\ {\rm km}$ neutron star, the recurrence time would be  $\approx 10$
years (we define $\dot m_{\rm Edd}=8.8\times 10^4\ {\rm g\ cm^{-2}\
s^{-1}}$, the local Eddington accretion rate for solar
composition). The pressure at the base of such a pile is
$P\approx 5.7\times 10^{26}(g_{14}/2)\ {\rm erg \ cm^{-3}}$, where the
surface gravity is $g=g_{14}10^{14}\ {\rm cm\ s^{-2}}$. At such high
pressures, the equation of state is that of degenerate, relativistic
electrons with Fermi energy $E_F=1.9 \ {\rm MeV}
(2Z/A)^{1/3}\rho_8^{1/3}$, where $\rho_8=\rho/10^8 \ {\rm g \
cm^{-3}}$.  Supporting this pressure requires a density $\rho\approx
1.3 \times 10^9 \ {\rm g \ cm^{-3}}$ and $E_{\rm F}\approx 4.3 \ {\rm
MeV}$.

Though they exert little pressure, the nuclei set the thermal
conductivity via electron-ion scattering.  For a classical one
component plasma with ion separation, $a$, defined by $a^3=3/4\pi
n_i$, where $n_i=\rho/A m_p$, the importance of Coulomb physics for
the ions is measured by
\begin{equation}
\label{eq:gamma}
\Gamma\equiv {(Ze)^2\over a kT} =434\ {\rho_8^{1/3}\over T_8}
\left(Z\over 44\right)^2\left(104\over A\right)^{1/3},
\end{equation}
where $T_8=T/10^8 {\rm K}$.  For this initial calculation we use the
microphysics of the liquid state; $\Gamma< 173$ (Farouki \& Hamaguchi
1993 and references therein), even when, in some instances (mostly
when $\dot m<0.3 \dot m_{\rm Edd}$) $\Gamma>173$. 

Electron conduction dominates the heat transport in the liquid
ocean. The heat flux is given by $F=-K dT/dz$, where $K$ is the
thermal conductivity (see Yakovlev \& Urpin 1980; Itoh et al.~1983,
Potekhin et al.~1999). For $X_{12}\ll 1$, we find
\begin{equation}
K\approx 1.1\times 10^{18}{\rm ergs \ cm^{-1} s^{-1}
K^{-1}}{\rho_8^{1/3}T_8\over Z^{2/3}A^{1/3}},
\label{eq:cond}
\end{equation}
where we assume the electrons are relativistic, and take the Coulomb
logarithm to be unity (a good approximation in the deep ocean;
Bildsten \& Cutler 1995). Integrating the heat equation and
hydrostatic balance gives the temperature contrast required between
two points (at pressure $P_2$ and $P_1$) to transport a heat flux $F$
(Bildsten \& Cutler 1995; Brown \& Bildsten 1998)
\begin{equation}
T_{8,2}^2\approx T_{8,1}^2+690{Z^2\over A}{F\over F_{\rm
Edd}}\ln\left(P_2\over P_1\right),
\label{eq:trel}
\end{equation} 
where $F_{\rm Edd}=\dot m_{\rm Edd}(GM/R)=8.8\times 10^{24}\ g_{14}\
{\rm erg\ cm^{-2}\ s^{-1}}$ is the Eddington flux for a $1.4\
M_\odot$, $10\ {\rm km}$ star.

The flux in the ocean is the sum of compressional heating and heat
flowing out from the crust, deposited there by electron captures and
pycnonuclear reactions (Haensel \& Zdunik 1990; Brown 2000). Since
both contributions are proportional to the local accretion rate per
unit area, we write $F=Q\dot m$, with $Q=10^{17}Q_{17} \ {\rm erg \
g^{-1}}$.  Equation (\ref{eq:trel}) then gives
\begin{equation}
T_{8,2}^2\approx T_{8,1}^2+{Q_{17}\over g_{14}}\left({\dot m\over\dot
m_{\rm Edd}}\right)\left({Z^2\over A}\right)\ln\left(P_2\over
P_1\right).
\label{eq:trel2}
\end{equation} 
In a careful study of the overall thermal balance of the neutron star,
Brown (2000) found $Q_{17}\approx 1$ for accretion rates $\gtrsim 0.1\
\dot m_{\rm Edd}$ (i.e., of the $\approx 1\ {\rm MeV}$ per nucleon
deposited in the crust, $\approx 10$\% escapes through the surface;
the rest is emitted as neutrinos from the inner crust and core).  The
compressional heating term contributes $\approx 11\ {\rm keV}\
(T_8/6)^2(1\ {\rm MeV}/E_F)(Z/44)(104/A)$ per nucleon, small compared
to the heat flux from the crust.

\begin{figure}
\epsscale{1.0}\plotone{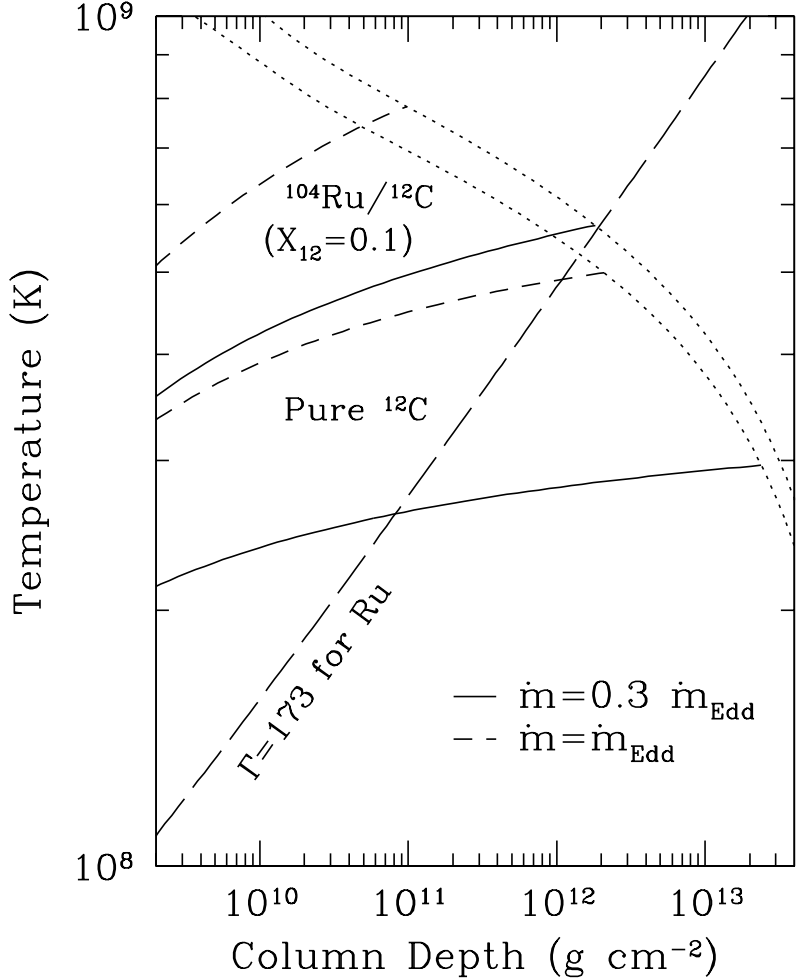}\epsscale{1.0} 
\figcaption{Thermal
profiles and ignition curves in the deep ocean of an accreting neutron
star. The ignition curves are shown as dotted lines for (bottom to
top) $X_{12}=1.0$ and $0.1$. The long-dashed curve shows where
$\Gamma=173$ for \Ru (the equivalent curve for \C lies off the right
of the plot). The temperature as a function of column depth for a
\C--\Ru mixture with $X_{12}=0.1$ is shown in the two upper curves,
while the lower two curves are for a pure \C ocean. We show two
accretion rates, $\dot m=0.3\ \dot m_{\rm Edd}$ (solid), and $\dot
m=\dot m_{\rm Edd}$ (dashed). We take $Q_{17}=1$. These profiles end at
the depth where \C ignites, as defined by equation (\ref{eq:ignite}).
\label{fig:settle}}
\end{figure}

For $\dot m\approx 0.1\ \dot m_{\rm Edd}$, the second term in equation
(\ref{eq:trel2}) is close to unity for a pure \C ocean ($Z^2/A=3$),
but $\approx 6$ times larger for a \Ru ocean ($Z^2/A=18.6$)
($\ln(P_2/P_1)\approx 8$). Thus, an ocean of heavy ashes is hotter
than that of pure \C at the same depth. This is critical, as it allows
for ignition at lower column densities, $y=P/g$. Figure
\ref{fig:settle} shows the temperature as a function of column depth
for a pure \C ocean, and for a \C--\Ru mixture with \C mass fraction
$X_{12}=0.1$. In each case, we take $Q_{17}=1$, and show two accretion
rates, $\dot m=0.3\ \dot m_{\rm Edd}$, and $\dot m=\dot m_{\rm
Edd}$. The dashed line is the crystallization condition for \Ru. The
lower thermal conductivity makes the \C--\Ru oceans much hotter than
the pure \C oceans.

We terminate the thermal profiles in Figure \ref{fig:settle} at the
depth at which the \C ignites. Brown \& Bildsten (1998) showed
that the condition for an unstable \C ignition in a neutron star ocean
is set by competition with conductive cooling, so that a 
thermal runaway occurs when the local
nuclear burning energy release, $\epsilon_{\rm nuc}$, satisfies
${d\epsilon_{\rm nuc}/d \ln T}>{d\epsilon_{\rm cool}/d\ln T}$, 
where $\epsilon_{\rm cool}=\rho KT/y^2$ is an approximation of the
cooling rate (e.g., Fujimoto, Hanawa, \& Miyaji 1981) and the nuclear
reaction rates are as described in Brown \& Bildsten (1998). For the thermal
conductivity given by equation (\ref{eq:cond}),
$d\epsilon_{\rm cool}/d \ln T= 2 \epsilon_{\rm cool}$, 
unstable ignition requires
\begin{equation}
\nu \epsilon_{\rm nuc}> {2\rho K T\over y^2}, 
\label{eq:ignite}
\end{equation}
where $\nu\equiv {d\ln \epsilon_{\rm nuc}/d \ln T}\approx 26$ is the
temperature sensitivity of the fusion reaction.
Ignition curves are shown in Figure \ref{fig:settle} for two
cases. The lower dotted line is for a pure \C ocean and matches that
in Brown \& Bildsten (1998). The upper dotted line shows the ignition
curve for a \C--\Ru mixture with $X_{12}=0.1$. Because of the lower
abundance, a larger column density is needed to ignite \C in a \Ru
ocean for a given temperature. However, ignition occurs at smaller
column depths in this case, because the \C--\Ru ocean is hotter. 

For the pure \C models of Figure \ref{fig:settle}, ignition occurs at
$y=2.1\times 10^{12}\ {\rm g\ cm^{-2}}$ ($y=2.4\times 10^{13}\ {\rm g\
cm^{-2}}$) for $\dot m=\dot m_{\rm Edd}$ ($\dot m=0.3\ \dot m_{\rm
Edd}$). Writing the burst energy $E_{\rm burst}=4\pi R^2yE_{\rm
nuc}X_{12}$, or $E_{\rm burst}=7\times 10^{42}\ {\rm erg}\
y_{12}X_{12}(R/10\ {\rm km})^2$, and recurrence time $t_{\rm
recur}=y/\dot m=0.36\ {\rm yr}\ y_{12}(\dot m_{\rm Edd}/\dot m)$, we
find that pure \C ignitions have recurrence times $0.8\ {\rm years}$
($29\
{\rm years}$) and burst energies $1.5\times 10^{43}\ {\rm ergs}$
($1.7\times 10^{44}\ {\rm ergs}$). These burst energies exceed the
energy of the superbursts by at least a factor of ten. For the \C--\Ru
models, ignition occurs at $y=1.0\times 10^{11}\ {\rm g\ cm^{-2}}$
($y=1.8\times 10^{12}\ {\rm g\ cm^{-2}}$) for $\dot m=\dot m_{\rm
Edd}$ ($\dot m=0.3\ \dot m_{\rm Edd}$), with recurrence time $13\ {\rm
days}$ ($2.2\ {\rm years}$), and burst energy $6.9\times 10^{40}\ {\rm
ergs}$ ($1.3\times 10^{42}\ {\rm ergs}$).

\section{Accretion Rate Dependences of the  Carbon Flashes} 

\begin{figure}
\epsscale{1.0}\plotone{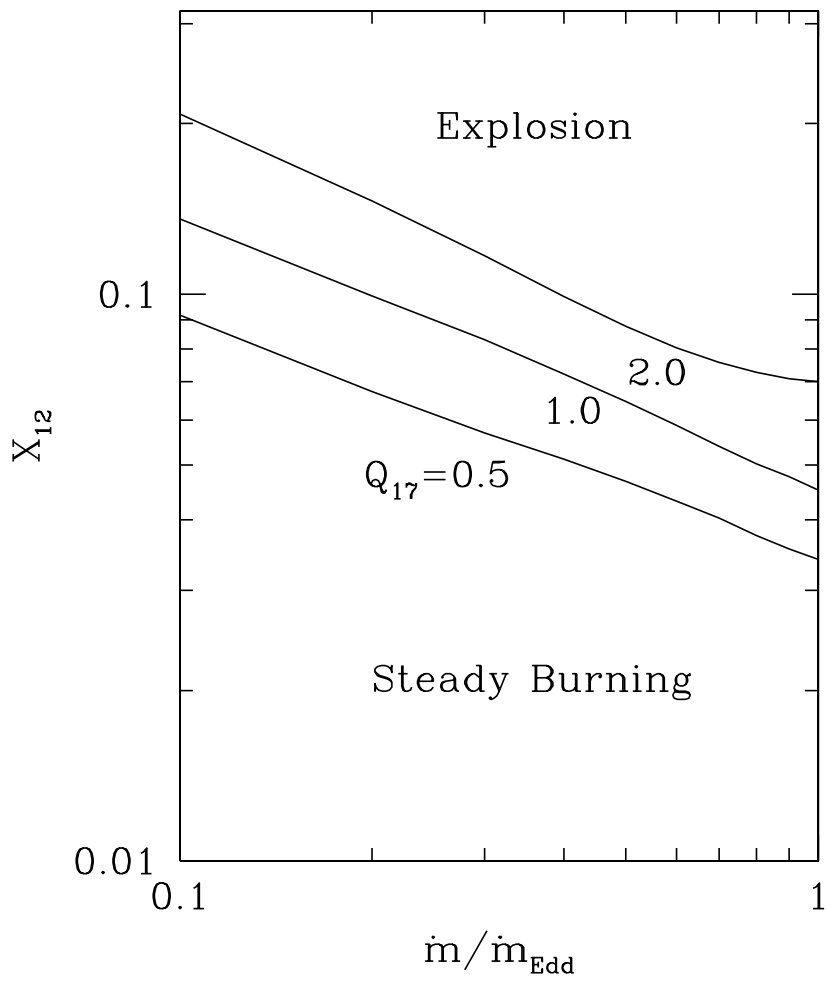}\epsscale{1.0} 
\figcaption{ The
minimum mass fraction of \C, $X_{12}$, needed to reach unstable
ignition before depletion, as a function of accretion rate for
$Q_{17}=0.5, 1.0$ and $2.0$ in a \Ru ocean. 
\label{fig:mdotX}}
\end{figure}

  Not only must the thermal settling solution reach the ignition
curve, but the \C must also have survived to that depth. Thus, in
order to obtain a flash, the \C lifetime to the fusion reaction,
$t_{\rm nuc}\approx E_{\rm nuc} X_{12}/\epsilon_{\rm nuc}$ must exceed
the accumulation time, $y/\dot m$, at the ignition point. Combining
this condition with equation (\ref{eq:ignite}) implies that 
$\dot m$ must exceed the critical value
\begin{equation}
\dot m_c\equiv {2\rho KT\over E_{\rm nuc} X_{12} y \nu}.
\end{equation} 
When we evaluate this using $K$ and the $\rho$, $y$ relations for
degenerate relativistic electrons, the density cancels out, giving
\begin{equation}\label{eq:mdotc}
{\dot m_c\over\dot m_{\rm Edd}}\approx\left({0.1\over X_{12}}\right)
\left({T_8\over 6}\right)^2
\left({26\over\nu}\right)\left({g_{14}\over 2}\right)
\left({A\over 104}\right)\left({44\over Z}\right)^2.
\end{equation}
This relation is robust and shows that large accretion rates are
required for unstable \C ignition. In order to eliminate the
temperature, we neglect the outer temperature in equation
(\ref{eq:trel2}) and substitute for the temperature at the base in
equation (\ref{eq:mdotc}). We then find that the explicit accretion
rate dependences cancel and the requirement for an instability is just
\begin{equation} 
\label{eq:XQ}
X_{12}>{4Q\over \nu E_{\rm nuc}}\ln\left(P_2\over
P_1\right)\approx 0.2\
Q_{17}\left({26\over\nu}\right){\ln\left(P_2/P_1\right)\over 8}. 
\end{equation}
At smaller $X_{12}$, the \C depletes and burns steadily before
reaching ignition. Figure \ref{fig:mdotX} shows the \C mass fraction
required to achieve an explosion as a function of $\dot m$  and
$Q_{17}$. The accretion rate dependence is mostly from
the ignition pressure dependence on $\dot m$. The top two panels in 
figure \ref{fig:gom} show the burst energies and recurrence times
for unstable models. The curves are for $X_{12}=0.1$ and (top
to bottom), $Q_{17}=0.5, 1.0$ and $2.0$. 

\section{Time Dependent Behavior of the Flash}

Having derived the conditions for an unstable ignition of a small mass
fraction of \C, we now calculate the time dependent nature of the
flash and estimate the time it takes for the energy to escape from the
burning layer.  The ``one-zone'' version of the time dependent heat
equation is
\begin{equation}
\label{eq:dtdt}
C_V {d T\over dt}=\epsilon_{\rm nuc}-\epsilon_{\rm cool},
\end{equation}
in which we write the specific heat at constant volume $C_V$ rather
than constant pressure (an excellent approximation since $E_F\gg k_BT$
throughout the flash). In the case of predominantly high $Z$ nuclei,
the electron specific heat dominates that of the ions, so
$C_V=(\pi^2Zk_B/Am_p)(k_BT/E_F)$.

Once ignited, the \C burning is extremely rapid and there is no time
for cooling to occur while the \C burns.  The fluid then
evolves to a final temperature set by the total energy released
$\int_{T_{\rm i}}^{T_{\rm f}} C_V dT=\int \epsilon_{\rm nuc} dt=X_{12}E_{\rm nuc}$. Using $C_V$
for the electrons and taking $T_{\rm f}\gg T_{\rm i}$ gives
\begin{equation}
\label{eq:tfinal}
T_{\rm f}=1.92\times 10^{9} {\rm K}\left(E_F\over {\rm MeV}\right)^{1/2}
\left(X_{12}\over 0.1\right)^{1/2}
\left(A\over 2.36Z\right)^{1/2}. 
\end{equation}
Thus, $X_{12}>0.01$ is adequate to substantially change the
temperature from that at ignition and trigger a thermal
instability. If $X_{12}>0.2-0.3$, the peak temperatures would
exceed $5\times 10^9 {\rm K}$ and neutrino cooling would reduce the
observed energetics of the event. 

\begin{figure}
\epsscale{1.0}\plotone{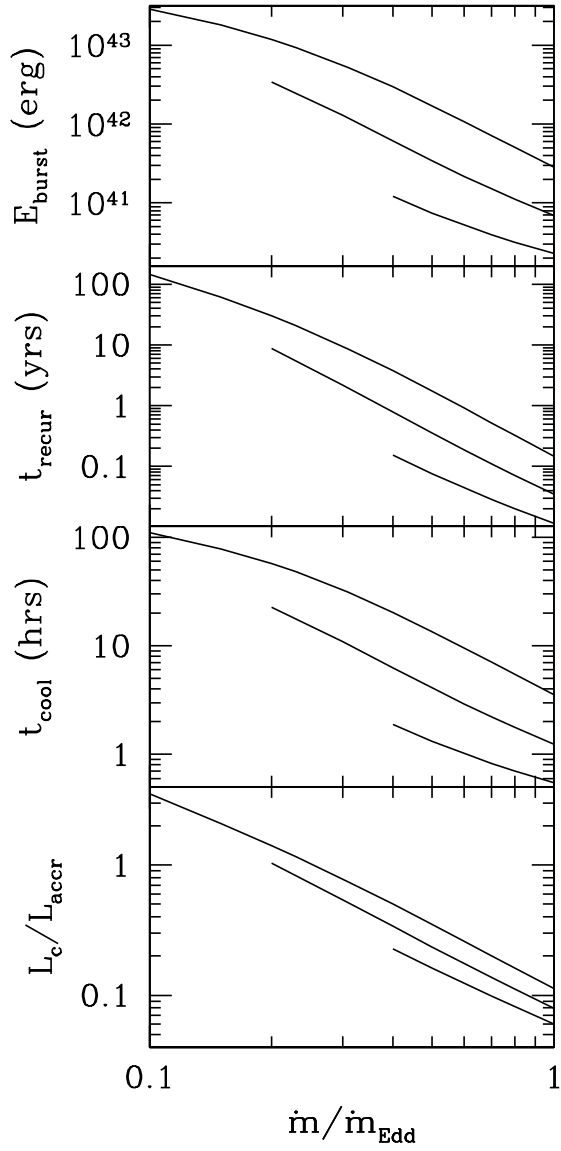}\epsscale{1.0} 
\figcaption{Burst energy,
recurrence time, luminosity decay time (equation [\ref{eq:tcool}]) and
$L_c$ (equation [\ref{eq:L}]) as a function of $\dot m$ for
$X_{12}=0.1$ in a \Ru ocean for (top to bottom) $Q_{17}=0.5, 1.0$ and
$2.0$. The curves begin at the lowest accretion rate at which the
thermal instability occurs in the \Ru ocean. 
\label{fig:gom}}
\end{figure}

After all the \C has burned and the fluid has reached $T_{\rm f}$, the
layer cools. Integrating equation (\ref{eq:dtdt}) starting from
$T_{\rm f}$ with only the $\epsilon_{\rm cool}$ term, we find $d\ln
T=-dt/2t_{\rm cool}$, where
\begin{equation}
\label{eq:tcool}
t_{\rm cool}=4.15\ {\rm hrs}\ {\rho_8\over g_{14}^2}
\left({Z\over 44}\right)^4\left({104\over A}\right)^3
\end{equation} 
depends only on the depth in the star and the composition, and is 
temperature independent. 
The luminosity during the conductive cooling phase
is thus set by the temperature at the base of the layer;
$L=4\pi R^2 \epsilon_{\rm cool} y\propto T^2\propto T_{\rm f}^2
\exp(-t/t_{\rm cool})$. Thus, the observer should see a timescale of
luminosity decay equal to equation (\ref{eq:tcool}), as shown in
Figure \ref{fig:gom}.  If we use $T_{\rm f}$ from equation
(\ref{eq:tfinal}) for an upper limit, we find the maximum conductive
cooling luminosity 
\begin{equation}\label{eq:L}
{L_c\over L_{\rm Edd}}=0.77\ X_{12}\rho_8^{1/3}\ \left({A\over
104}\right)^{5/3}\left({44\over Z}\right)^{8/3}. 
\end{equation}
Due to our neglect of convective heat transport, this $L_c$ will be
less than the peak luminosity of the burst.  However, since convection
cannot last much longer than the burning time, we expect something similar to
this simple luminosity evolution to occur during the burst decay, 
over times relevant to detection by flux monitoring on hourly timescales.

Detection of these flashes should be easiest at low $\dot M$'s, as
$L_c$ increases due to ignition at higher densities, whereas the
accretion luminosity, $L_{\rm accr}=GM\dot M/R$, is dropping. Hence,
the contrast of the burst cooling tail is largest for low $\dot M$
sources, as shown in the bottom panel of Figure \ref{fig:gom}.
Detecting one of these flares from a rapidly accreting $Z$ source
would require flux sensitivity at the $10\%$ level on a timescale of a
few hours and spectral sensitivity to distinguish that the flux rise
is from extra thermal emission. Since these bursts will occur roughly
once a month, it is an archival project to find the rare event in the
accumulated observations.

\section{Conclusions and Future Work}

We have shown that the ignition of a small mass fraction of \C in a
neutron star ocean consisting mostly of heavier elements is thermally
unstable and leads to flashes whose energy, recurrence times and
durations are similar to the superbursts. The energy from these
flashes takes a long time to escape the star, possibly explaining the
persistent ``offset'' in the flux nearly a day after the superburst in
4U 1735-44 (see Figure 1 of Cornellise et al. 2000). We also predict
flashes at high local accretion rates which include the Z sources and
accreting X-ray pulsars. In the pulsar case, some of the difficulties
associated with explaining flares from LMC X-4 via pure \C burning
(see Brown \& Bildsten 1998) might well be alleviated in the mixed
case we present here.

Cornellise et al. (2000) reported an absence of regular type I bursts
for seven days after the superburst. This might be due to thermal
stabilization of the hydrogen/helium burning layers by the large heat
flux from the cooling ashes of the \C burning.  Paczynski (1983) and
Bildsten (1995) have shown that luminosities in excess of the helium
burning flux (or $L>L_{\rm accr}/100$) will stabilize the burning, as
then the thermal state of the burning layer is independent of the
local burning rates. This appears most likely at low accretion rates,
where the conductive cooling is slowest, thus halting Type I bursts
for $\approx 5 t_{\rm cool}\sim$ days after the superburst.

There is still much to be done, including a time dependent study of
the onset of the instability.  Our initial integrations of the
time-dependent code from Bildsten (1995) demonstrate a transition from
steady-state burning at low $\dot m$ to unstable flashes at higher
rates. In future work, this code will be used to more accurately
calculate the transition accretion rate, $\dot m_c$ with realistic
mixes of \C and rp-process ashes. Only then can we make the
comparisons to observations that will allow us to constrain the
rp-process ashes.

\acknowledgements 
We thank Erik Kuulkers and Marten van Kerkwijk for
conversations about the properties of the superbursts and Hendrik
Schatz for insights on the rp-process. We also thank Deepto
Chakrabarty and Bob Rutledge for comments on the manuscript. This
research was supported by NASA via grant NAG 5-8658 and by the
National Science Foundation under Grants PHY99-07949 and AY97-31632.
L. B. is a Cottrell Scholar of the Research Corporation.


\end{document}